\documentclass[final,5p,twocolumn]{elsarticle}

\usepackage{lineno}

\usepackage{amsmath,amssymb,bm,subfigure,color,url,booktabs,mathrsfs}

\definecolor{darkblue}{RGB}{0,50,90}

\usepackage[colorlinks=true,citecolor=blue,linkcolor=blue,urlcolor=blue]{hyperref}

\graphicspath{{fig/}}

\journal{Information Sciences}

\begin{document}

\begin{frontmatter}



\title{
Non-Local Musical Statistics as Guides for Audio-to-Score Piano Transcription\tnoteref{label1}
}
\tnotetext[label1]{This work was in part supported by JSPS KAKENHI Nos.\ 16H01744 and 19K20340, JST ACCEL No.\ JPMJAC1602, and the Kyoto University Foundation.}

\author[alabel1]{Kentaro Shibata}
\author[alabel1,alabel2]{Eita Nakamura\corref{cor1}}
\ead{eita.nakamura@i.kyoto-u.ac.jp}
\author[alabel1]{Kazuyoshi Yoshii}
\cortext[cor1]{Corresponding author}
\address[alabel1]{Graduate School of Informatics, Kyoto University, Kyoto 606-8501, Japan}
\address[alabel2]{The Hakubi Center for Advanced Research, Kyoto University, Kyoto 606-8501, Japan}

\begin{abstract}
We present an automatic piano transcription system that converts polyphonic audio recordings into musical scores. This has been a long-standing problem of music information processing, and recent studies have made remarkable progress in the two main component techniques: multipitch detection and rhythm quantization. Given this situation, we study a method integrating deep-neural-network-based multipitch detection and statistical-model-based rhythm quantization. In the first part, we conducted systematic evaluations and found that while the present method achieved high transcription accuracies at the note level, some global characteristics of music, such as tempo scale, metre (time signature), and bar line positions, were often incorrectly estimated. In the second part, we formulated non-local statistics of pitch and rhythmic contents that are derived from musical knowledge and studied their effects in inferring those global characteristics. We found that these statistics are markedly effective for improving the transcription results and that their optimal combination includes statistics obtained from separated hand parts. The integrated method had an overall transcription error rate of $7.1\%$ and a downbeat F-measure of $85.6\%$ on a dataset of popular piano music, and the generated transcriptions can be partially used for music performance and assisting human transcribers, thus demonstrating the potential for practical applications.
\end{abstract} 




\end{frontmatter}


\section{Introduction}
\label{sec:Intro}

Automatic music transcription has been a long-standing fundamental problem in music informatics \cite{Benetos2019}.
The ultimate goal is to convert music audio signals into musical scores, which are useful for music performance and music content analysis.
For example, a vast number of music audio and video files are available on the Web, and for most of them, it is difficult to find the corresponding musical scores, which are necessary for practicing music, making covers, and conducting detailed music analysis.
The central problem of automatic music transcription is to obtain symbolic representation of musical pitches and rhythms from continuous signals.
Transcribing polyphonic music, which contains multiple pitches sounding simultaneously, is especially a challenging problem because of the huge search space and the difficulty of separating individual pitches from a sound mixture; it is difficult even for human experts.
Here we study the problem of transcribing polyphonic piano music, which is one of the major forms of music.

Due to the complexity of the problem, polyphonic music transcription has been studied as two-split problems, multipitch detection and rhythm quantization.
In multipitch detection, an audio signal is converted into a {\it performance MIDI sequence}, which is a list of musical notes with semitone-level pitches, onset and offset times in seconds, and velocities (intensities).
Spectrogram factorization methods such as nonnegative matrix factorization (NMF) and probabilistic latent component analysis (PLCA) have been common approaches to this problem \cite{Benetos2015,Cheng2016,Vincent2010}.
More recently, significant improvements have been achieved by means of deep neural network (DNN) techniques \cite{Bittner2017,Hawthorne2018,Sigtia2016,Wu2019}.

In rhythm quantization, a performance MIDI sequence is converted into a {\it quantized MIDI sequence} where the onset and offset times are described in units of beats.
In this task, utilizing musical knowledge about tempo changes and common rhythmic patterns is essential and methods based on statistical models such as hidden Markov models (HMMs) have been studied for recognizing quantized onset times \cite{Cemgil2000,Hamanaka2003,Nakamura2017,Raphael2002}.
For recognizing quantized offset times, or equivalently note values, a method based on Markov random field has been proposed \cite{Nakamura2017B}.

Despite the active research in these two fields, studies on the whole audio-to-score transcription problem are still scarce \cite{Kapanci2005}.
As a recent attempt, \cite{Nakamura2018ICASSP} proposed an audio-to-score piano transcription system that integrates a multipitch detection method based on PLCA and a rhythm quantization method based on HMM.
That paper concluded that the results were often far from the practical level because of the limited performance of the multipitch detection method.
The system proposed in \cite{Cogliati2018} uses a multipitch detection method based on convolutional sparse coding and a MIDI-to-score conversion method \cite{Cogliati2016} that uses the Melisma Analyzer \cite{Temperley2009} for rhythm quantization.
No accounts of the full audio-to-score transcription system, however, have been reported in the literature.
Another direction of research is the end-to-end approach to audio-to-score transcription \cite{Carvalho2017,Nishikimi2019,Roman2019}.
At present, however, the reported studies cover only constrained conditions (e.g.\ synthetic sound) and are of limited success.
Given the significant progress of DNN-based multipitch detection methods, currently the most promising approach is to integrate one of these methods with the best-performing rhythm quantization method.

Most recent studies on piano transcription rely on the MAPS data \cite{MAPS} for evaluation.
This dataset consists mostly of Western classical music, which is considered to be a reasonable source of experimental data for its variety and complexity and for the lack of concerns over copyright issues.
However, musical scores of classical music are easily accessible and there are few demands for new transcriptions.
From a practical viewpoint, much more commercial and academic demands are expected in the field of popular music.
Since popular music and classical music have different features, it is important to evaluate a transcription system with popular music data to examine its potential and limitations in a practical situation.

This study is composed of two parts.
The purpose of the first study is to examine the potential of the integration of DNN-based multipitch detection and statistical-model-based rhythm quantization methods.
We explicitly construct an audio-to-score (wav to MusicXML) piano transcription system and conduct systematic evaluations using classical music and popular music data.
As a result, we found that although the system achieves high performance in terms of note-level evaluation metrics, it makes a significant amount of errors for global musical characteristics; the most common errors are misidentification of tempo scale (halved tempos), metre (confusion of 4/4 time and 3/4 time), and positions of bar lines (downbeats).
The result indicates that these global characteristics cannot be accurately inferred from local musical statistics considered in the applied statistical models.
As time signature and bar lines are pieces of basic information for understanding the structure of music, it is crucial for applications that they are correctly given in transcribed scores.

Given these results, the purpose of the second study is to understand the principles for correctly estimating these global musical characteristics.
In cognitive music theory, it has been argued that various musical features are involved in the recognition and representation of metrical structure \cite{GTTM}.
Studies on metre detection \cite{Gainza2009}, beat tracking \cite{Dixon2000}, and musical structure analysis \cite{Nieto2020} have also suggested the importance of non-local features, such as self-similarity and voice configuration, for determining musical characteristics related to metrical structure.
Gathering this knowledge from several research fields, we formulate a set of musical statistics and conduct experiments to find out the relevance of each statistic and the optimal combination of statistics for improving the transcribed results.
The results indicate that non-local statistics are useful guides for inferring the global characteristics and that a specific combination of statistics has a significantly better effect than using all the statistics or using only local ones.

Compared to the previous systems \cite{Nakamura2018ICASSP,Roman2019}, the present method achieved a considerable improvement and approach towards a practical audio-to-score music transcription system.
As examples in the accompanying webpage\footnote{\url{https://audio2score.github.io/}} demonstrate, transcribed scores can partly be used in practice and can assist human transcribers.
We also discuss current limitations and implications for further studies on automatic music transcription.

\section{Method for Audio-to-Score Piano Transcription}
\label{sec:System}

\subsection{System Architecture}

%
\begin{figure}
\centering
\includegraphics[width=0.5\columnwidth]{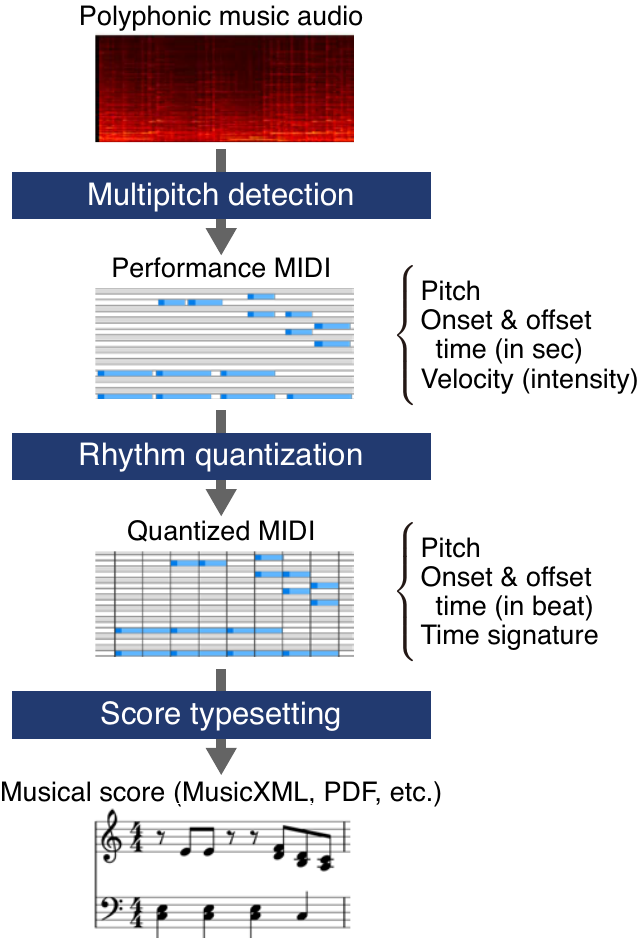}
\caption{Outline of the piano transcription system.}
\label{fig:SystemArchitecture}
\end{figure}

The outline of the present audio-to-score piano transcription system is shown in Fig.~\ref{fig:SystemArchitecture}.
In the multipitch detection step, a performance MIDI sequence is estimated for an input audio signal.
In the rhythm quantization step, the onset and offset times in the performance MIDI sequence are quantized and represented in beat units.
In the score typesetting step, the quantized MIDI sequence is converted to a MusicXML file, which is a common data format for human/computer-readable score representation.
We explain these three steps in the following sections.

\subsection{Multipitch Detection}
\label{sec:MultipitchDetection}

\begin{figure}
\centering
\includegraphics[width=1\columnwidth]{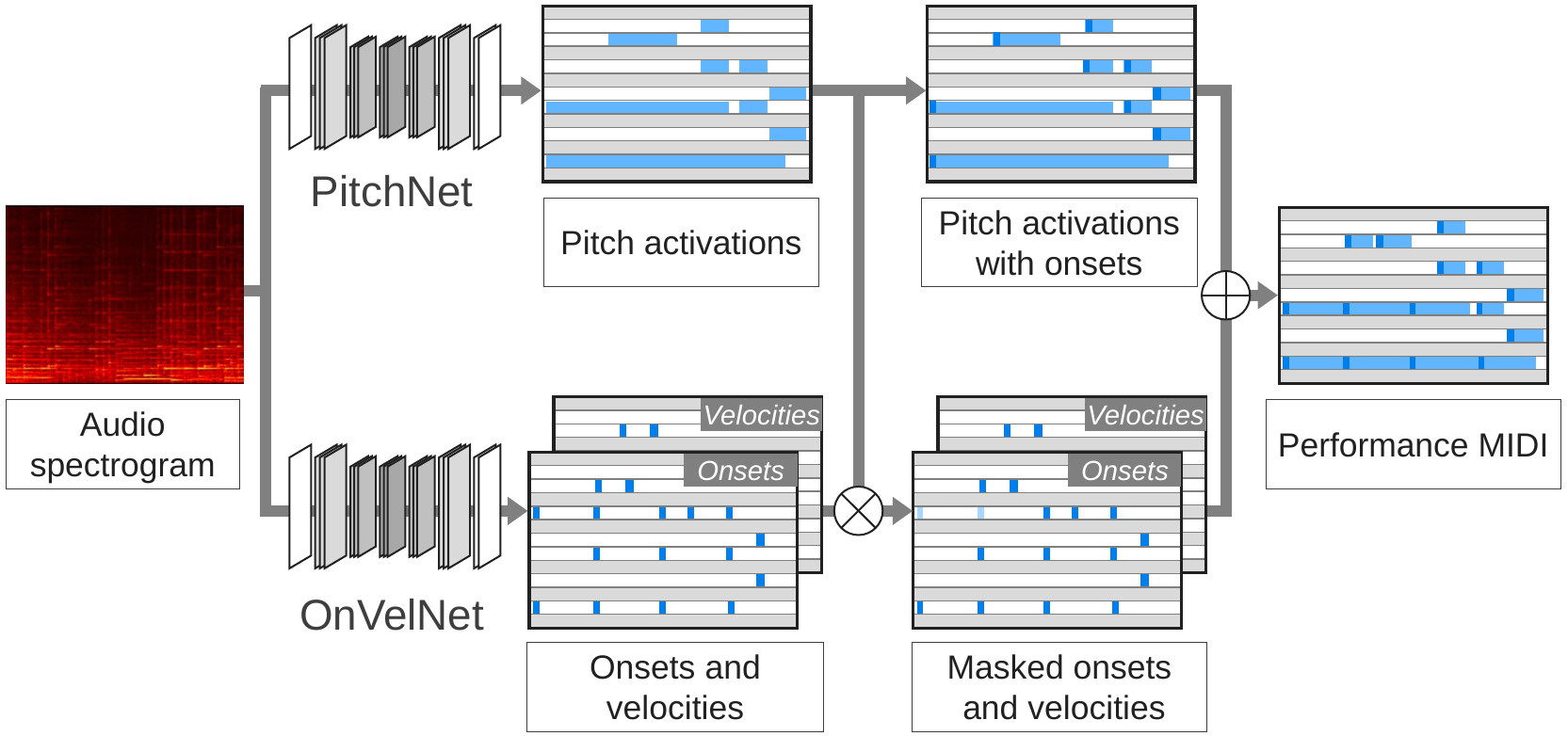}
\caption{Architecture of the multipitch detection method (POVNet).}
\label{fig:POVNet}
\end{figure}

We use a convolutional neural network (CNN) called \textit{DeepLabv3+}, which was first used for image segmentation \cite{Chen2018} and later applied for multipitch detection \cite{Wu2019}.
The original network \cite{Wu2019} estimates only pitch activation and we modify it to estimate onset and velocity as well.
The multipitch detection method (POVNet) consists of two DNNs, one for pitch analysis (PitchNet) and the other for onset and velocity estimation (OnVelNet) (Fig.~\ref{fig:POVNet}).
These networks are trained separately, and a performance MIDI sequence is obtained by combining their outputs.
The inputs to these networks are harmonic combined frequency and periodicity (HCFP) features \cite{Wu2019} denoted by $\bm Z \in \mathbb{R}_+^{2H \times F \times T}$, where $H$ is the number of harmonic partials, $F$ the number of frequency bins, and $T$ the number of time frames.

Given HCFP features ${\bm Z}$ as input, PitchNet outputs an $F\times T$ probability matrix ${\bm P}_{\rm p}$, whose element ${\bm P}_{\rm p}(f,t)\in [0,1]$ represents the salience of frequency $f$ at frame $t$.
The network architecture is the same as in \cite{Wu2019} ($F=352=88\times4$ and $H=6$). 
In the last layer, ${\bm P}_{\rm p}$ is obtained by a sigmoid function.
PitchNet is trained by a binary cross-entropy loss function
\begin{align}\label{eq:pitch_loss}
{\cal L}_{\rm p}
= -\frac{1}{FT} \sum_{f,t=1}^{F,T}&\bigg[\big\{1-\widehat{\bm P}_{\rm p}(f,t)\big\}{\rm ln}\big\{1-{\bm P}_{\rm p}(f,t)\big\}
\notag\\
&\quad+\widehat{\bm P}_{\rm p}(f,t)\,{\rm ln}\,{\bm P}_{\rm p}(f,t)\bigg],
\end{align}
where $\widehat{\bm P}_{\rm p} \in \{0,1\}^{F \times T}$ denotes a binary pitch activation matrix obtained from the ground-truth MIDI data (sustain pedal events are taken into account).
Finally, an $M\times T$ pitch activation matrix ${\bm D}_{\rm p}$, whose element ${\bm D}_{\rm p}(m,t)\in\{0,1\}$ represents the presence of semitone-level pitch $m$ at frame $t$, is obtained by binarizing and down-sampling ${\bm P}_{\rm p}$ along the frequency axis ($M=88$ is the number of pitches on a piano keyboard).

Given HCFP features $\bm Z$ as input,
OnVelNet outputs an onset probability matrix ${\bm P}_{\rm o} \in [0,1]^{F \times T}$
and an intensity matrix ${\bm P}_{\rm v} \in [0,1]^{F \times T}$, whose elements ${\bm P}_{\rm o}(f,t)$ and ${\bm P}_{\rm v}(f,t)$ represent the onset salience and the intensity, respectively, at frequency $f$ and frame $t$.
The intensity here takes a real value between $0$ and $1$ corresponding to an integral value between $0$ and $127$ defined as velocity in the MIDI format.
OnVelNet has the same architecture as PitchNet except for the last layer, where ${\bm P}_{\rm o}$ is obtained by a sigmoid function and ${\bm P}_{\rm v}$ is obtained by a clip function with an interval of $[0,1]$.
This network is trained by minimizing the weighted sum ${\cal L}_{\rm ov}=w_{\rm o}{\cal L}_{\rm o}+ w_{\rm v}{\cal L}_{\rm v}$ of a binary cross-entropy loss $ {\cal L}_{\rm o}$ and a mean squared error loss ${\cal L}_{\rm v}$ given by
\begin{align}
\label{eq:onvel_loss}
{\cal L}_{\rm o} &= - \frac{1}{FT}\sum_{f,t=1}^{F,T}\bigg[
\big\{1-\widehat{\bm P}_{\rm o}(f,t)\big\}\,{\rm ln}\big\{1-{\bm P}_{\rm o}(f,t)\big\}
\notag\\
&\hspace{70pt}+\widehat{\bm P}_{\rm o}(f,t)\,{\rm ln}\,{\bm P}_{\rm o}(f,t)\bigg],
\\
{\cal L}_{\rm v} &= \frac{1}{FT}\sum_{f,t=1}^{F,T}\widehat{\bm P}_{\rm o}(f,t)\left\{ \widehat{\bm P}_{\rm v}(f,t) - {\bm P}_{\rm v}(f,t)\right\}^{2}.
\end{align}
Here, $\widehat{\bm P}_{\rm o} \in \{0,1\}^{F \times T}$ is a binary matrix representing the presence of note onsets and $\widehat{\bm P}_{\rm v} \in [0,1]^{F \times T}$ is a real-valued matrix representing the intensities of note onsets.
These two matrices are obtained from the ground-truth MIDI data.
To allow small fluctuations of onset times, $\widehat{\bm P}_{\rm o}(f,t \pm 1)$ are set to $1$ if $\widehat{\bm P}_{\rm o}(f,t)=1$ originally.
Finally, an onset matrix ${\bm D}_{\rm o} \in \{0,1\}^{M \times T}$ is obtained by binarizing ${\bm P}_{\rm o}$ and down-sampling the result along the frequency axis.
If ${\bm D}_{\rm o}$ has successive $1$\,s along the time axis, these elements are set to $0$ except the $1$ at the centre.
A velocity matrix ${\bm D}_{\rm v} \in \{0,\ldots,127\}^{M \times T}$ is obtained by applying scaling, rounding-down, and down-sampling to ${\bm P}_{\rm v}$.

A performance MIDI sequence $(t_n,\bar{t}_n,p_n,v_n)^N_{n=1}$ representing the onset times $t_n$, offset times $\bar{t}_n$, pitches $p_n$, and velocities $v_n$ of notes is obtained from pitch activations ${\bm D}_{\rm p}$, onset matrix ${\bm D}_{\rm o}$, and velocity matrix ${\bm D}_{\rm v}$ ($N$ is the number of notes in the MIDI sequence).
To ensure the consistency between the pitch activations and onsets, ${\bm D}_{\rm o}(m,t)$ is set to $0$ if ${\bm D}_{\rm p}(m, t)=0$.
Onset times are obtained by picking time-frequency bins that satisfy ${\bm D}_{\rm p}(m, t)=1$ and ${\bm D}_{\rm p}(m, t{-}1)=0$, and ${\bm D}_{\rm o}$ is used for detecting successive notes.
The following rules are applied as well:
\begin{itemize}\setlength\itemsep{-0.2em}\setlength{\itemindent}{-0pt}
\item If an onset from ${\bm D}_{\rm p}$ and one from ${\bm D}_{\rm o}$ are within $100$ ms, they are merged by retaining only the earlier one.
\item Notes with durations of $<30$ ms are removed.
\item Notes with velocities of $<40$ are removed.
\end{itemize}
The threshold on the durations is set to eliminate unphysical notes ($30$ ms approximately corresponds to a 64th note in a tempo of 120 beats per minutes (BPM)).
The threshold on the velocities has the effect of reducing false positives with some risk of missing soft notes.
The value is not tuned and was determined after some trials.

\subsection{Rhythm Quantization}
\label{sec:RhythmQuantization}

Given a performance MIDI sequence $(t_n,\bar{t}_n,p_n,v_n)^N_{n=1}$, the rhythm quantization method estimates a quantized MIDI sequence $(\tau_n,\bar{\tau}_n,p_n)^N_{n=1}$.
The onset score times $\tau_n$ are estimated first (onset rhythm quantization step) and the offset score times $\bar{\tau}_n$ are estimated subsequently (note value recognition step).
For onset rhythm quantization, we use the metrical HMM \cite{Hamanaka2003,Raphael2002} extended for polyphonic music \cite{Nakamura2018ICASSP}.
For note value recognition, we use the method in \cite{Nakamura2017B}.
As preparation for later discussions, we here summarize the onset rhythm quantization method.

The metrical HMM describes the generative process of onset score times, local tempos, and onset times.
Onset score times $\tau_n$ are generated by a Markov model with initial probability $P(\tau_1)$ and transition probabilities $P(\tau_n|\tau_{n-1})$.
These probabilities are represented in terms of metrical positions $b_n$, which indicate the positions of $\tau_n$ relative to bar lines.
Onset score times and metrical positions are described in units of tatums (minimal resolution of beats).
The tatum unit is assumed to be $1/3$ of a 16th note in this study.
The length $B$ of a bar is determined by the underlying metre (for example, $B=48$ for $4/4$ time and $B=36$ for $3/4$ time) and metrical position $b_n$ has a value in $\{0,1,\ldots,B-1\}$ where $b_n=0$ indicates the downbeat (beginning of a bar).
In addition, we introduce a chord variable $g_n$ that indicates whether the $(n-1)$th and $n$\,th notes have the same onset score time ($g_n={\rm CH}$) or not ($g_n={\rm NC})$.
Based on this data representation, the initial probability is given as $P(\tau_1)=P(b_1)$ and the transition probabilities are given as
\begin{align}
P(\tau_n|\tau_{n-1})=\chi_{b_{n-1},g_n}(\delta_{g_n,{\rm CH}}\delta_{b_n,b_{n-1}}+\delta_{g_n,{\rm NC}}\pi_{b_{n-1},b_n}),
\label{eq:ScoreModel}
\end{align}
where $\delta$ is Kronecker's symbol, $\chi_{b_{n-1},g_n}=P(g_n|b_{n-1})$ are the probabilities of chordal notes at each metrical position, and $\pi_{b_{n-1},b_n}=P(b_n|b_{n-1})$ are the metrical transition probabilities.
The difference $\tau_n-\tau_{n-1}$ between consecutive onset score times is determined as
\begin{align}
\tau_n-\tau_{n-1}=\begin{cases}
0,&g_n={\rm CH};\\
b_n-b_{n-1},&g_n={\rm NC},b_n>b_{n-1};\\
b_n-b_{n-1}+B,&g_n={\rm NC},b_n\leq b_{n-1}.
\end{cases}
\end{align}
The chord probabilities $\chi_{b_{n-1},g_n}$ and metrical transition probabilities $\pi_{b_{n-1},b_n}$ describe the frequencies of rhythmic patterns used in music and are learned from musical score data.

The local tempos $u_n$ describe the ratio of the onset time scale described in seconds and the score time scale described in tatum units.
To allow tempo variations, they are assumed to obey a Gaussian-Markov model:
\begin{align}
u_1={\rm Gauss}(u_{\rm ini},\sigma^2_{{\rm ini}\,u}),\quad u_n={\rm Gauss}(u_{n-1},\sigma^2_u).
\label{eq:TempoModel}
\end{align}
Here, ${\rm Gauss}(\mu,\sigma^2)$ denotes a Gaussian distribution with mean $\mu$ and standard deviation $\sigma$, $u_{\rm ini}$ represents the average initial tempo, $\sigma_{{\rm ini}\,u}$ the amount of global tempo variation, and $\sigma_u$ the amount of local tempo changes.
Given the sequence of onset score times $\tau_n$ and local tempos $u_n$, the onset times $t_n$ are generated by the Gaussian/Exponential model in \cite{Nakamura2015} as
\begin{align}
P(t_n)=\begin{cases}
{\rm Gauss}(t_{n-1}+u_{n-1}(\tau_n-\tau_{n-1}),\sigma^2_t),&g_n={\rm NC};
\\
{\rm Exp}(t_{n-1},\lambda_t),&g_n={\rm CH},
\end{cases}
\label{eq:TimingModel}
\end{align}
where ${\rm Exp}(x,\lambda)$ denotes an exponential distribution with scale parameter $\lambda$ and support $[x,\infty)$.
The parameters $\sigma_t$ and $\lambda_t$ represent the fluctuation of onset times; the former is for time intervals between chords and the latter for asynchrony of chordal notes.

Putting together the probabilistic models in Eqs.~(\ref{eq:ScoreModel}), (\ref{eq:TempoModel}), and (\ref{eq:TimingModel}), we can calculate the joint probability $P(t_{1:N},\tau_{1:N},u_{1:N})$ ($t_{1:N}$ denotes $(t_n)_{n=1}^N$ etc.).
Given the onset times $t_{1:N}$, we can estimate the onset score times $\tau_{1:N}$ and local tempos $u_{1:N}$ by maximizing the probability $P(\tau_{1:N},u_{1:N}\,|t_{1:N})\propto P(t_{1:N},\tau_{1:N},u_{1:N})$.
This can be done by the Viterbi algorithm with discretization of the tempo variables \cite{Nakamura2018ICASSP}.

So far, we have assumed that the underlying metre and the corresponding bar length $B$ are given.
To estimate the metre of the input performance, we can apply the maximum likelihood method \cite{Nakamura2017}.
The procedure is as follows: we construct multiple metrical HMMs corresponding to candidate metres ($4/4$, $3/4$, $2/4$, etc.), calculate the maximum probability $P(t_{1:N},\tau_{1:N},u_{1:N})$ for each model, and finally obtain the most probable metre according to the probability.

\subsection{Score Typesetting}

To convert a quantized MIDI sequence to graphical musical score representation, it is necessary to properly assign the musical notes to either the right-hand part or the left-hand part.
To do this, we use the hand part separation method of \cite{Nakamura2014}.
There are often more than one melody (or ``voice'' in musical terminology) in each hand part, and in that case, it is necessary to separate the voices as well.
Although several voice separation methods exist \cite{Cambouropoulos2008,DeValk2018,McLeod2016}, some assume strictly monophonic voices, which is inappropriate for general piano music, and the others have not been made available for public use.
Therefore, we implemented a cost-function-based voice separation method that can handle homophonic voices.
Since we need some space to describe the method in detail and it is not the main scope of this study, the voice separation method is presented in \ref{app:VoiceSeparation}.
The result of hand part and voice separation is preserved as a quantized MIDI sequence with separated tracks, each corresponding to voices.
In the last step of score typesetting, we use the public software MuseScore 3\footnote{MuseScore 3, \url{https://musescore.org/}.} to obtain score notation in the MusicXML format.
To properly preserve the voice structure, the quantized MIDI sequence is imported to MuseScore and a MusicXML file with multiple voice tracks is exported.
The final transcription result in the MusicXML format is obtained by merging the voice tracks in each hand part into a single staff.

\section{Systematic Evaluation}
\label{sec:Evaluation}

\subsection{Data and Experimental Setups}

We use two kinds of music data, classical music and popular music, for evaluating the transcription system.
To simplify the experimental setup, we train and test the methods separately for these two sets of data.
The purpose of using classical music data is to enable comparison with existing methods and we use the conventionally used MAPS dataset \cite{MAPS}.
This dataset contains piano pieces by various composers and the audio recordings of MIDI sequences with manually added musical expressions.
Specifically, for testing, we use the 60 pieces labelled ``ENSTDkCl'' and ``ENSTDkAm,'' which are recordings of MIDI piano playbacks of the MIDI data.
For training the chord probabilities and metrical transition probabilities of the metrical HMM, the dataset of classical music in \cite{Nakamura2017B} is used, as is done in \cite{Nakamura2018ICASSP}.
We use the same parameterization for the performance model as in \cite{Nakamura2018ICASSP}: the tempo variables are discretized into 50 values logarithmically equally spaced in the range between $u_{\rm min}=0.3$ s/QN (sec per quarter note) and $u_{\rm max}=1.5$ s/QN (corresponding to BPM 40 and BPM 200), $\sigma_u=3.32\times 10^{-2}$ s/QN, $u_{\rm ini}=\sqrt{u_{\rm max}u_{\rm min}}$, $\sigma_{{\rm ini}\,u}=3\sigma_u$, $\sigma_t=0.02$ s, and $\lambda_t=0.0101$ s.
We also use the default parameter values for the note value recognition method as in \cite{Nakamura2018ICASSP}.

The purpose of using popular music data is to examine the system's performance in a practical situation, as discussed in the Introduction.
For testing, we collected 81 piano covers of popular music whose audio recordings are available on YouTube and corresponding musical scores are available from a public website\footnote{Yamaha Music Entertainment Holdings, Print gakufu, \url{https://www.print-gakufu.com/}.}.
These pieces were selected from the most popular pieces and were intended to cover a variety of artists and different levels of performance difficulty.
The pieces were played by various pianists; some were played with a real piano and the others were played with a digital piano.
The quality of the audio recordings was generally high.
Since most pieces are J-pop songs, we hereafter call this dataset the {\it J-pop dataset}.
For training the chord probabilities and metrical transition probabilities of the metrical HMM for popular music, we used a collection of 811 pieces, which were obtained from a public website\footnote{MuseScore 3, \url{https://musescore.org/}.}.
We downloaded all musical scores that appeared by searching `piano cover’ in the website and removed noisy ones with obviously irregular typesetting.
We call this dataset the {\it MuseScore dataset}.
For the parameters of the performance model, we use a parameterization slightly different from the one for classical music because the amount of tempo changes is usually smaller in popular music performances.
We set $\sigma_u=3.32\times 10^{-3}$ s/QN and $\sigma_t=0.03$ s.
We use the default parameter values for the note value recognition method as in the case of classical music.

The POVNet was trained with the MAPS dataset excluding the ``ENSTDkCl'' and ``ENSTDkAm'' subsets by using the RAdam optimizer~\cite{Liu2019} with a standard initial learning rate of $0.001$.
We set the loss weights $w_{\rm o}=0.9$ and $w_{\rm v}=0.1$, taking into account the importance of onset detection and the difficulty of velocity estimation.
The frame shift for the HCFP features was $20$ ms and inputs to the CNNs had $512$ frames ($10.24$ s) and were shifted by $128$ frames ($2.56$ s).

\subsection{Accuracy of Multipitch Detection}

We first evaluated the performance of the multipitch detection method, since it is an important component to compare with the previous method \cite{Nakamura2018ICASSP}.
For this purpose, we use the MAPS dataset, which includes the ground-truth MIDI data, and the frame-level metrics and note-level metrics defined in \cite{Bay2009}; the dataset and metrics are conventionally used in the research field.
In the frame-level metrics, the precision ${\cal P}_{\rm f}$, recall ${\cal R}_{\rm f}$, and F-measure ${\cal F}_{\rm f}$ are calculated with a time resolution of $10$ ms.
In the note-level metrics, the precision ${\cal P}_{\rm n}$, recall ${\cal R}_{\rm n}$, and F-measure ${\cal F}_{\rm n}$ are calculated by judging detected onsets that are within $\pm 50$ ms from ground-truth onsets as correct.
For consistency with previous studies, we used the first $30$ s of each recording.

\begin{table}[t]
\centering
\tabcolsep = 4pt
\begin{tabular}{lcccccc}
\toprule
Method & ${\cal P}_{\mathrm{f}}$ & ${\cal R}_{\mathrm{f}}$ & ${\cal F}_{\mathrm{f}}$ 
 & ${\cal P}_{\mathrm{n}}$ & ${\cal R}_{\mathrm{n}}$ & ${\cal F}_{\mathrm{n}}$\\
\midrule
PLCA \cite{Nakamura2018ICASSP} &
--- & --- & --- & 77.9 & 68.9 & 72.8
\\
OaF \cite{Hawthorne2019} &
{\bf92.9} & 78.5 & 84.9 & 87.5 & {\bf 85.6} &  {\bf 86.4}
\\ 
DeepLabv3+ \cite{Wu2019} &
87.5 & {\bf86.3} & {\bf 86.7} & --- & --- & ---
\\
PitchNet only &
89.3 & 84.4 & {\bf 86.6} & {\bf 91.1} & 68.4 & 77.5
\\
POVNet &
89.3 & {\bf 85.7} & {\bf 87.3} & 89.7 & 84.1 & {\bf 86.7}
\\ 
\bottomrule
\end{tabular}
\caption{Accuracies (\%) of multipitch detection on the MAPS-ENSTDkCl and MAPS-ENSTDkAm datasets. The best values (within a range of $1$ percentage point (PP)) are indicated in bold font.}
\label{tab:frame_eval}
\vspace{-5pt}
\end{table}

The results are summarized in Table \ref{tab:frame_eval}.
In addition to POVNet, the PLCA method used in \cite{Nakamura2018ICASSP}, a representative DNN-based method \cite{Hawthorne2019} (OaF; {\it Onsets and Frames} trained with the MAESTRO dataset), the original DeepLabv3+ in \cite{Wu2019}, and the results using only PitchNet are compared in the table.
POVNet outperformed the others in both the frame-level and note-level F-measures.
POVNet and OaF had equivalent ${\cal F}_{\rm n}$, which were significantly higher than ${\cal F}_{\rm n}$ for the PLCA method.
The difference in ${\cal R}_{\rm n}$ between POVNet and the method using only PitchNet clearly demonstrates the efficacy of OnVelNet, which enabled detection of repeated tones.

\subsection{Accuracy of Audio-to-Score Transcription}
\label{sec:AccuracyOfTranscription}

To evaluate the performance of audio-to-score transcription systems, we use the edit-distance-based metrics defined in~\cite{Nakamura2018ICASSP} and the MV2H metrics defined in~\cite{McLeod2018}.
In the former metrics, the following error rates (ERs) are calculated: pitch ER ${\cal E}_{\rm p}$, missing note rate ${\cal E}_{\rm m}$, extra note rate ${\cal E}_{\rm e}$, onset time ER ${\cal E}_{\rm on}$, offset time ER ${\cal E}_{\rm off}$, and overall (average) ER ${\cal E}_{\rm all}$.
MV2H calculates accuracies/F-measures of multipitch detection ${\cal F}_{\rm p}$, voice separation ${\cal F}_{\rm voi}$, metrical alignment ${\cal F}_{\rm met}$, note value detection ${\cal F}_{\rm val}$, and harmonic analysis ${\cal F}_{\rm harm}$, as well as the average of them ${\cal F}_{\rm MV2H}$.
${\cal F}_{\rm met}$ measures the correctness of beat assignment in levels of bar, beat, and sub-beat.
${\cal F}_{\rm harm}$ is in general a weighted sum of the chord accuracy and key accuracy, but only the key accuracy is used here because the tested methods do not estimate chord labels.

For evaluation on classical music data, we used the 30 pieces in the MAPS-ENSTDkCl dataset as in \cite{Nakamura2018ICASSP}.
For the onset rhythm quantization method, three metrical HMMs corresponding to bar lengths of 4 quarter notes ($4/4$ time), 3 quarter notes ($3/4$ time and $6/8$ time), and 2 quarter notes ($2/4$ time) were constructed, and the metre was estimated by the method described in Sec.~\ref{sec:RhythmQuantization}.
For comparison, we applied the same rhythm quantization method to the performance MIDI sequences obtained by the PLCA method \cite{Nakamura2018ICASSP} and to the ground-truth MIDI data.
For evaluation on popular music data, where most pieces have either 4/4 time or 3/4 time, two metrical HMMs corresponding to bar lengths of 4 quarter notes and 3 quarter notes were constructed, and metre was estimated similarly.
We also tested the onset rhythm quantization method trained with the classical music data in this case to examine the effect of using music data of different genres for training.

\begin{table}[t]
\centering
\tabcolsep = 3pt
\begin{tabular}{lcccccc}
\toprule
Perform.\ MIDI & ${\cal E}_{\rm p}$ & ${\cal E}_{\rm m}$ &${\cal E}_{\rm e}$ & ${\cal E}_{\rm on}$ & ${\cal E}_{\rm off}$ & ${\cal E}_{\rm all}$
\\
\midrule
PLCA \cite{Nakamura2018ICASSP} & 4.96 & 25.7 & 16.4 & 28.3 & 41.6 & 23.4
\\
POVNet & {\bf 1.24} & {\bf 7.90} & {\bf 6.02} & {\bf 11.9} & {\bf 28.1} &{\bf 11.0}
\\
\midrule
Ground truth & 1.03 & 2.07 & 2.33 & 4.63 & 21.08 & 6.23
\\
\midrule
CTD16$^*$ \cite{Cogliati2016} & 1.12 & 13.6 & 6.49 & 17.1 & 44.0 & 16.5
\\
\bottomrule
\toprule
Perform.\ MIDI & ${\cal F}_{\rm p}$ & ${\cal F}_{\rm voi}$ & ${\cal F}_{\rm met}$ & ${\cal F}_{\rm val}$ & ${\cal F}_{\rm harm}$ & ${\cal F}_{\rm MV2H}$
\\
\midrule
PLCA \cite{Nakamura2018ICASSP} & 67.4 & 65.3 & 30.0 & 82.8 & 58.7 & 60.8
\\
POVNet & {\bf 85.0} & {\bf 67.5} & {\bf 41.4} & {\bf 87.3} & {\bf 71.7} & {\bf 70.6}
\\
\midrule
Ground truth & 91.2 & 71.1 & 51.7 & 91.3 & 77.0 & 76.5
\\
\midrule
CTD16$^*$ \cite{Cogliati2016} & 81.0 & 53.3 & 42.4 & 85.2 & 72.7 & 66.9
\\
\bottomrule
\end{tabular}
\caption{Error rates (\%) and accuracies (\%) of transcription on the MAPS-ENSTDkCl dataset. For comparison of the PLCA method and POVNet, a better value is indicated in bold font if the difference is larger than 1 PP. The POVNet's MIDI outputs were used as the inputs to the CTD16 method~\cite{Cogliati2016}. $^*$Calculated from the 27 (out of 30) pieces for which the system could output results.}
\label{tab:score_eval_classical}
\end{table}

The results for the classical music data are shown in Table \ref{tab:score_eval_classical}.
The system using POVNet outperformed the system using the PLCA method in all metrics.
In particular, large decreases in the edit-distance-based error rates were observed, which clearly confirms the significant effect of using the improved multipitch detection method.
Among the edit-distance-based metrics, the onset time ER and offset time ER were still relatively high for the POVNet-based system, indicating the difficulty of precisely recognizing rhythms.
Among the MV2H metrics, the metrical accuracy, which also measures the accuracy of transcribed rhythms, was particularly low.
The fact that a variety of metres are used in the classical music data also made it difficult for the method to correctly estimate metres.
The result for the ground-truth MIDI data shows that further improvements are expected by refining the multipitch detection method, the note value recognition method, and the voice separation method.

The results for the popular music data are shown in Table \ref{tab:score_eval_popular}.
Overall, the error rates were lower and accuracies were higher compared to the case of MAPS data, indicating that the difficulty of transcription is generally lower for the popular music data.
Notably, for these data, the voice and metre accuracies were around $80\%$.
This is because piano pieces of popular music genre usually have simple voice structure (melody in the right-hand part and chord accompaniment in the left-hand part) and simple metrical structure ($96\%$ of the pieces are in 4/4 time and the others are in 3/4 or 6/8 time).
As for the effect of using music data of different genres for training the onset rhythm quantization method, significant improvements were observed for the onset time ER and the metrical accuracy by using training data of the same genre as the test data.

\begin{table}[t]
\centering
\tabcolsep = 3pt
\begin{tabular}{lcccccc}
\toprule
Training data & ${\cal E}_{\rm p}$ & ${\cal E}_{\rm m}$ &${\cal E}_{\rm e}$ & ${\cal E}_{\rm on}$ & ${\cal E}_{\rm off}$ &${\cal E}_{\rm all}$
\\
\midrule
Classical music & 0.59 & 4.12 & 7.38 & 3.62 & 21.0 & 7.35
\\
MuseScore & 0.62 & 4.09 & 7.35 & {\bf 2.50} & 20.8 & 7.06
\\
\midrule
CTD16$^*$ \cite{Cogliati2016} & 1.51 & 12.8 & 7.27 & 9.25 & 55.3 & 17.2
\\
\bottomrule
\toprule
Training data & ${\cal F}_{\rm p}$ & ${\cal F}_{\rm voi}$ & ${\cal F}_{\rm met}$ & ${\cal F}_{\rm val}$ & ${\cal F}_{\rm harm}$  & ${\cal F}_{\rm MV2H}$
\\
\midrule
Classical music & 93.2 & 79.4 & 63.7 & 92.9 & 91.9 & 84.2
\\
MuseScore & 93.2 & 79.4 & {\bf 80.3} & {\bf 95.2} & 92.0 & {\bf 88.0}
\\
\midrule
CTD16$^*$ \cite{Cogliati2016} & 86.3 & 42.3 & 58.2 & 82.8 & 91.2 & 72.2
\\
\bottomrule
\end{tabular}
\caption{Error rates (\%) and accuracies (\%) of transcription on the J-pop dataset. The training data indicate that used for the metrical HMM for rhythm quantization. Performance MIDIs obtained by POVNet were used. For comparison of the training datasets, a better value is indicated in bold font if the difference is larger than 1 PP. $^*$Calculated from the 72 (out of 81) pieces for which the system could output results.}
\label{tab:score_eval_popular}
\vspace{-5pt}
\end{table}
\begin{table*}[t]
\centering
\tabcolsep = 5pt
\begin{tabular}{cccccccccc}
\toprule
Test data & Method & ${\cal A}_{\rm metre}$ & ${\cal A}_{\rm tempo}$ & ${\cal P}_{\rm B}$ & ${\cal R}_{\rm B}$ & ${\cal F}_{\rm B}$ & ${\cal P}_{\rm DB}$ & ${\cal R}_{\rm DB}$ & ${\cal F}_{\rm DB}$\\
\midrule
MAPS & MetHMM & 23.3 & 50.0 & {\bf75.7} & 76.7 & 73.6 & 45.5 & {\bf42.6} & {\bf42.2} \\
& CTD16$^*$ \cite{Cogliati2016} & 25.9 & --- & 73.3 & {\bf93.9} & {\bf79.6} & {\bf48.7} & {\bf41.9} & {\bf42.9} \\
& LPCFG \cite{McLeod2017} & {\bf50.0} & --- & 73.0 & 58.3 & 62.4 & 35.1 & 35.3 & 32.1 \\
\midrule
J-pop & MetHMM & {\bf87.7} & 76.5 & {\bf95.1} & {\bf87.1} & {\bf89.8} & {\bf74.9} & {\bf67.1} & {\bf69.4} \\
& CTD16$^*$ \cite{Cogliati2016} & 62.5 & --- & 84.2 & 85.8 & 83.8 & 45.7 & 38.5 & 40.9 \\
& LPCFG \cite{McLeod2017} & 64.2 & --- & 86.7 & 71.7 & 77.0 & 53.8 & 45.0 & 47.2 \\
\bottomrule
\end{tabular}
\caption{Accuracies of metrical structure estimated by the metrical HMM (MetHMM), the CTD method \cite{Cogliati2016}, and the lexicalized probabilistic context-free grammar (LPCFG) model \cite{McLeod2017}. The best values (within a range of 1 PP) are indicated in bold font. $^*$Calculated from the pieces for which the system could output results (see the captions to Tables \ref{tab:score_eval_classical} and \ref{tab:score_eval_popular}).}
\label{tab:AccuracyOfMetricalStructure}
\vspace{-5pt}
\end{table*}

We also evaluated the MIDI-to-score conversion method proposed by Cogliati et al.~\cite{Cogliati2016}, which is a component of the audio-to-score transcription system proposed in \cite{Cogliati2018}.
Since the released source code\footnote{\url{https://github.com/AndreaCogliati/CompleteTranscription}} could not output musical scores in the MusicXML format, we extracted from the program the information necessary for score typesetting and used MuseScore 3 to obtain transcription results in the MusicXML format.
The CTD16 method uses the rhythm quantization and voice separation functions of the Melisma Analyzer version 2 \cite{Temperley2009}, and for the evaluation metrics considered here, the results largely reflect the ability of this analyser.
The default settings were used for the Melisma Analyzer.

The results are shown in Tables \ref{tab:score_eval_classical} and \ref{tab:score_eval_popular}, where the method by Cogliati et al.~\cite{Cogliati2016} is represented as CTD16.
Outputs could not be obtained for some pieces due to runtime errors of the Melisma Analyzer and those pieces were excluded from the calculation of the evaluation metrics.
It is notable that the missing note rates were significantly higher compared to the present method.
We confirmed that this was due to the behaviour of the Melisma Analyzer; its output often contained fewer notes than the input.
The onset time ER and offset time ER were also significantly higher, which is consistent with the results in previous studies \cite{Nakamura2017B,Nakamura2017}.
Among the MV2H metrics, the CTD16 method had significantly lower voice accuracies for both the MAPS and J-pop datasets.
This result reflected the limitation of the Melisma Analyzer, that it can only estimate monophonic voices.
For the classical music data, the metrical accuracy and harmonic accuracy for the CTD16 method were each higher than those for the present method, but the differences were small (within 1 percentage point (PP)).
These results demonstrate the strength of the present {\it learning-based} statistical method for rhythm quantization compared to the CTD16 method (or the Melisma Analyzer) whose parameters were manually adjusted.

\subsection{Analysis of Errors Regarding Metrical Structure}
\label{sec:ErrorAnalysis}

To investigate tendencies of transcription errors regarding metrical structure in more detail, we additionally use the following metrics.
We define the metre accuracy ${\cal A}_{\rm metre}$ as the proportion of musical pieces for which the transcribed score has the same bar length as the ground truth (the most frequent metre was taken as ground truth in case of a piece with mixed metres).
The tempo scale accuracy ${\cal A}_{\rm tempo}$ is defined as the proportion of pieces for which the estimated global tempo $\bar{u}_{\rm est}$ and the global tempo $\bar{u}_{\rm true}$ of the ground-truth musical score satisfy a condition $0.8\,\bar{u}_{\rm true}\leq \bar{u}_{\rm est}\leq 1.2\,\bar{u}_{\rm true}$.
To measure the accuracy of beat estimation, the beat precision ${\cal P}_{\rm B}$, recall ${\cal R}_{\rm B}$, and F-measure ${\cal F}_{\rm B}$ are defined.
When both a note in the ground-truth score and the corresponding note in the transcription have an onset on a beat, the transcribed note is counted as a true positive.
Similarly, the downbeat precision ${\cal P}_{\rm DB}$, recall ${\cal R}_{\rm DB}$, and F-measure ${\cal F}_{\rm DB}$ are defined.

The results are shown in Table \ref{tab:AccuracyOfMetricalStructure}, where the rhythm quantization method is indicated as `MetHMM'.
In the case of classical music (MAPS data), the metre accuracy and downbeat F-measure were especially low, which are consequences of the variety of time signatures used.
In the more concerning case of popular music (J-pop data), the accuracies were high overall, but the tempo scale accuracy and downbeat F-measures were low.
Given that $95\%$ of the pieces in this dataset are in 4/4 time, the relatively low metre accuracy indicates that the metrical HMM is not close to perfect for discriminating between 4/4 time and 3/4 time.
We found that most of the incorrectly estimated tempo scales had halved tempos compared to the ground truth, which was the cause for the low beat and downbeat recalls.
We thus conclude that estimation errors in tempo scale, metre, and downbeat positions are common ones regarding the metrical structure.

In Table \ref{tab:AccuracyOfMetricalStructure}, the results obtained by applying the CTD16 method \cite{Cogliati2016} and the metrical alignment method based on a lexicalized probabilistic context-free grammar (LPCFG) \cite{McLeod2017}, instead of the metrical HMM, are shown for comparison.
The latter method is one of the state-of-the-art methods for metre detection and downbeat estimation for symbolic music data.
The accuracies of tempo scales are not shown for these methods because they do not explicitly estimate them.
For the classical music data, although the LPCFG method had the highest metre accuracy, it had lower beat and downbeat F-measures than the other methods.
The CTD16 method had a very high beat recall, which led to a higher beat F measure than the metrical HMM.
A cause for the gap in the beat recall values is that the 8-beat times (6/8 and 9/8) were not incorporated in the metrical HMM.
Another possible reason is that the CTD16 method (or the Melisma Analyzer) takes into account harmony and pitch features, which are not incorporated in the metrical HMM.
We also found that the Melisma Analyzer tends to output more notes on beats, which led to the high beat recall: $68\%$ of notes were on beats in the transcription results by the method, whereas $48\%$ of notes were on beats in the ground-truth data (the results by the metrical HMM had $49\%$ of notes on beats).
The metrical HMM and the CTD16 method had similar downbeat F-measures.

For the popular music data, the metrical HMM outperformed the other methods for all the metrics, often by large margins.
This is possibly because the LPCFG method was trained on classical music data and the default parameters for the Melisma Analyzer were not suited for popular music.
From these results, it is confirmed that correctly estimating the tempo scale, metre, and downbeat positions is still difficult for the existing methods for metrical structure analysis, particularly when applied for automatically transcribed scores.

\section{Non-Local Musical Statistics}
\label{sec:NonLocalStatistics}

%
\begin{table}[t]
\centering
\tabcolsep = 3pt
\begin{tabular}{lll}
\toprule
Sec. & Symbol & Meaning
\\
\midrule
\ref{sec:MultipitchDetection} & $n$ & Musical note index
\\
& $t_n$ ($\bar{t}_n$) & Onset (offset) time
\\
& $p_n$ & Pitch
\\
\midrule
\ref{sec:RhythmQuantization} & $\tau_n$ ($\bar{\tau}_n$) & Onset (offset) score time
\\
& $b_n$ & Metrical position
\\
\midrule
\ref{sec:AccuracyOfTranscription} & ${\cal E}_{\rm p}$, ${\cal E}_{\rm m}$, etc. & Edit-distance-based metrics
\\
& ${\cal F}_{\rm p}$, ${\cal F}_{\rm voi}$, etc. & MV2H metrics
\\
\midrule
\ref{sec:ErrorAnalysis} & ${\cal A}_{\rm metre}$ & Metre accuracy
\\
& ${\cal A}_{\rm tempo}$ & Tempo scale accuracy
\\
& ${\cal P}_{\rm B}$, ${\cal R}_{\rm B}$, ${\cal F}_{\rm B}$ & Metrics for beat estimation
\\
& ${\cal P}_{\rm DB}$, ${\cal R}_{\rm DB}$, ${\cal F}_{\rm DB}$ & \begin{tabular}{@{\hspace{0em}}l}Metrics for downbeat\\estimation\end{tabular}
\\
\midrule
\ref{sec:NonLocalStatistics} & $A_4$, $A_3$ & Auto-similarity indices
\\
& $L^{\rm BH}_{\rm met}$, $L^{\rm RH}_{\rm met}$, $L^{\rm LH}_{\rm met}$ & Log metrical prob.
\\
& $L^{\rm BH}_{\rm NV}$, $L^{\rm RH}_{\rm NV}$, $L^{\rm LH}_{\rm NV}$ & Log note value prob.
\\
& $R^{\rm BH}_{\rm tie}$, $R^{\rm RH}_{\rm tie}$, $R^{\rm LH}_{\rm tie}$ & Negative rate of ties
\\
& $L^{\rm BH}_{\rm rel.pc}$, $L^{\rm RH}_{\rm rel.pc}$, $L^{\rm LH}_{\rm rel.pc}$ & \begin{tabular}{@{\hspace{0em}}l}Log prob.\ of relative\\pitch classes\end{tabular}
\\
& $C^{\rm BH}_{\rm SSM}$, $C^{\rm RH}_{\rm SSM}$, $C^{\rm LH}_{\rm SSM}$ & SSM contrast index
\\
& $L_{\rm p.rank}$ & Log prob.\ of pitch ranks
\\
\bottomrule
\end{tabular}
\caption{List of frequently used symbols and non-local statistics. The left column indicates the sections where the symbols are introduced.}
\label{tab:MathSymbols}
\end{table}

As discussed in the previous section, common errors made by the automatic transcription method are related to the tempo scale, metre, and positions of bar lines (downbeats).
According to our musical knowledge, these global characteristics cannot be completely inferred from local statistics that are considered in the metrical HMM or similar generative models.
We here formulate several musical statistics that are motivated by musical knowledge and expected to play a role in recognizing the global characteristics.

First, since it is possible to rescale the tempo and correspondingly the beat unit without changing musical interpretation, the tempo scale is intrinsically arbitrary, and convention plays an essential role in its choice.
For example, metres such as $3/8$ time and $3/2$ time were common in the Baroque period, but they are rarely used in contemporary popular music.
Therefore, the {\it mean tempo} and the {\it mean note value} within a piece are basic statistics whose distributions reflect the convention.

Second, metrical structure is related to repetitions in multiple scales (bar, phrase, period, section, etc.) \cite{GTTM}.
It is thus natural to consider autocorrelation \cite{Brown1993} or self-similarity \cite{Gainza2009} of musical elements to induce the metre of a musical sequence.
We formulate the beat-level self-similarity matrix for a musical score $X$ as follows.
Recall the mathematical notations listed in Table \ref{tab:MathSymbols}.
For convenience, we index beats $i$ as $i=0,1,\ldots,I-1$ where $I$ is the length of $X$ in beat units.
We use $X_i$ to represent the set of indices of notes contained in the musical score segment between beat $i$ and beat $i+\Delta$ ($\Delta$ is the window size).
We introduce a similarity measure $D(X_i,X_j)$ for two segments $X_i$ and $X_j$; $D(X_i,X_j)$ is assumed to take values between $0$ and $1$, and a larger value indicates higher similarity.
Based on the musical knowledge that repetitions in music can involve the pitch content, the rhythmic content, or both, we formulate the similarity measure as
\begin{align}
D(X_i,X_j)&=\frac{D_{\rm p}(X_i,X_j)+D_{\rm r}(X_i,X_j)}{2},
\\
D_{\rm p}(X_i,X_j)&=\frac{2|{\rm Pitch}(X_i)\cap{\rm Pitch}(X_j)|}{|{\rm Pitch}(X_i)|+|{\rm Pitch}(X_j)|},
\\
D_{\rm r}(X_i,X_j)&=\frac{2|{\rm NV}(X_i)\cap{\rm NV}(X_j)|}{|{\rm NV}(X_i)|+|{\rm NV}(X_j)|}.
\end{align}
Here, ${\rm Pitch}(X_i)=\{(\tau_n,p_n)|n\in X_i\}$ denotes the pitch content of segment $X_i$, whose elements are indicated by a pair of score time $\tau_n$ and pitch $p_n$, ${\rm NV}(X_i)=\{(\tau_n,r_n)|n\in X_i\}$ denotes the note-value content of segment $X_i$, whose elements are indicated by a pair of score time $\tau_n$ and note value (score-notated duration) $r_n=\bar{\tau}_n-\tau_n$, and $|S|$ denotes the cardinality of a set $S$.
It is straightforward to check $0\leq D(X_i,X_j)\leq 1$ for any segments $X_i$ and $X_j$, and $D(X_i,X_i)=1$ unless $X_i$ is empty (we define $D(X_i,\phi)=D(\phi,X_i)=0$ for an empty set $\phi$).
We call $D_{ij}=D(X_i,X_j)$ the {\it self-similarity matrix} (SSM).

\begin{figure*}
\centering
\includegraphics[width=1.5\columnwidth]{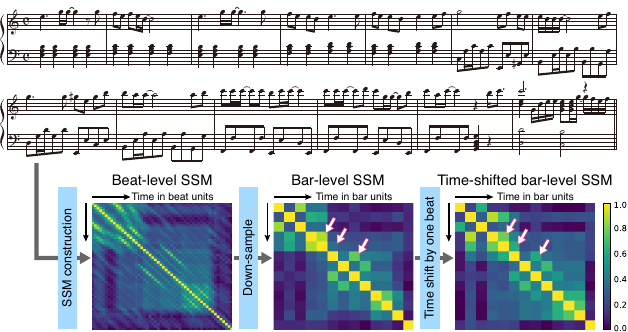}
\caption{Example of a self-similarity matrix (SSM) of a piano score (Piece No.~55 from the RWC popular music database \cite{RWC} arranged for piano). In the bar-level SSMs, arrows indicate elements that have extreme values in the original SSM but not in the time-shifted one.}
\label{fig:SSM}
\end{figure*}

We now define the {\it auto-similarity function} $A(X;s)$ of a musical score $X$ (with segments $\{X_i\}$) as
\begin{align}
A(X;s)=\frac{1}{I-s}\sum^{I-s-1}_{i=0}D(X_i,X_{i+s}),
\end{align}
where $s$ is time lag.
Since repetitions (including approximate one) usually occur in units of bars, we expect a large value of $A(X;s)$ if $s$ is a multiple of the bar length of $X$.
In the application to the transcription of popular music, a bar length of 4 beats (4/4 time) and that of 3 beats (3/4 time and 6/8 time) are of utmost importance.
Thus, we define the {\it auto-similarity index} of period 4 $A_4$ and that of period 3 $A_3$ as
\begin{align}
A_4&=\frac{1}{4}\{A(X;4)+A(X;8)+A(X;12)+A(X;16)\},
\\
A_3&=\frac{1}{4}\{A(X;3)+A(X;6)+A(X;9)+A(X;12)\}.
\end{align}
An example of an SSM computed from a piano score in 4/4 time is shown in Fig.~\ref{fig:SSM}.
Line-shaped patterns parallel to the diagonal line indicate repeated segments.
We can observe that the distances of these line-shaped patterns from the diagonal line are mostly multiples of 4 beats, reflecting that repetitions occur in units of bars.
These patterns contribute to $A_4$.

Third, whereas the metre is related to the period of repetitions in music, the bar line positions are related to their phase.
Therefore, it is essential to look for features that differ significantly when musical scores are tentatively shifted in time in beat units.
Since metrical structure is related to multiple aspects of music \cite{GTTM}, there are several statistics with this property.
The {\it log metrical probability} $L_{\rm met}=\sum_n{\rm ln}\,P(\tau_n|\tau_{n-1})$ represents the likelihood of the configuration of onsets, where $P(\tau_n|\tau_{n-1})$ is given by Eq.~(\ref{eq:ScoreModel}).
It is the statistic used to determine downbeat positions by the metrical HMM.
It is known that notes on downbeat positions tend to have longer durations \cite{Dixon2000}, which suggests the use of the {\it log note value (NV) probability} $L_{\rm NV}$.
This statistic is formulated as the likelihood of the configuration of note values given onset metrical positions and mathematically given as $L_{\rm NV}=\sum_n{\rm ln}\,P(r_n|b_n)$ where $r_n$ denotes the note value of the $n$\,th note and $b_n$ denotes its metrical position, both in tatum units.
A simpler quantity to represent a particular aspect of note values is the {\it negative rate of ties across a bar} $R_{\rm tie}$, which is defined as the ratio of the number of ties across a bar and the total number of notes, multiplied by $-1$.
Since we expect fewer ties across a bar for musical scores with correct bar lines (metrical preference rule (MPR) 8 in \cite{GTTM}), we define a negative quantity to conform with other quantities that tend to have a maximal value for correctly positioned downbeats.

For tonal music with which we are concerned, the tonal structure tends to align with the metrical structure.
One such property is that chord changes, especially cadences, tend to occur at strong beats (\cite{GTTM}, MPR 7).
Statistically, this can be formulated as the {\it log probability of relative pitch classes} (rel.pc) $L_{\rm rel.pc}=\sum_n{\rm ln}\,P(q_n|b_n)$ defined conditionally on metrical positions.
Here, $q_n\in\{0,1,\ldots,11\}$ is the pitch class of the $n$\,th note relative to the tonic of the local key ($q_n=0$ indicates a tonic tone, $q_n=7$ a dominant tone, etc.).
Another property is that bass notes tend to be positioned at strong beats (\cite{GTTM}, MPR 6).
As bass notes are characterized by locally lowest pitches, we define the {\it log probability of pitch ranks} $L_{\rm p.rank}=\sum_n{\rm ln}\,P(e_n|b_n)$, where the {\it pitch rank} (p.rank) $e_n$ of note $n$ denotes the rank of its pitch $p_n$ among the $K$ nearest pitches, i.e.\ $e_n={\rm Rank}(p_n,p_{n+1},\ldots,p_{n+K-1})$.

In \cite{GTTM}, it is pointed out that the boundaries of musical sections are usually drawn at bar lines and are often indicated by changes in musical features.
For example, accompaniment patterns and rhythmic patterns of melodies often change at section boundaries; the piano score in Fig.~\ref{fig:SSM} is a typical example of this.
On the other hand, more repetitions of musical features tend to be found within each section.
In the context of musical structure analysis, the first property is called novelty and the second one is called homogeneity, and both of them are used as useful guides to detect section boundaries \cite{Nieto2020}.
This indicates that phrases and sections are often recognized as block diagonal patterns in SSMs, as seen in the example of Fig.~\ref{fig:SSM}.
Therefore, when the SSM of a musical score is down-sampled at downbeat units, its nearly diagonal elements tend to have values distributed around end points $1$ and $0$ for correctly positioned downbeats and the distribution becomes less acute if the musical score is tentatively shifted in time.
In Fig.~\ref{fig:SSM}, the bar-level SSM for the original piece and that for the same piece, but all notes are time shifted in one beat are shown.
The latter SSM has less contrasting elements (indicated by arrows) and overall looks more like a blurred image.
Based on this observation, we formulate the {\it SSM contrast index} $C_{\rm SSM}$ as
\begin{align}
C_{\rm SSM}&=\sum^{J-2}_{k=0}\frac{C(D_{kM,(k+1)M})+C(D_{kM,(k+2)M})}{2(J-1)},
\end{align}
where $M$ is a (prospective) bar length, $J=\lfloor (I-1)/M\rfloor$ is the corresponding number of bar lines, and the {\it contrast function} $C(x)$ is defined as
\begin{align}
C(x)&=(x-1/2)^2-1/4.
\end{align}
This function has maxima $0$ at $x=0$ and $x=1$, and a minimum $-1/4$ at $x=1/2$ so that the index $C_{\rm SSM}$ has a larger value for an SSM with higher contrast; the last constant $-1/4$ is introduced to eliminate the influence of empty bars (whole rests).
There are other functions that satisfy these conditions and we chose the quadratic function here for mathematical simplicity.
We set the SSM window length $\Delta=M$ for computing this quantity.

Lastly, since vocal melodies and instrumental accompaniments have different characteristics, it is considered relevant to formulate the  statistics introduced here separately for right- and left-hand parts.
We use exactly the same formulation for a musical score $X^{\rm RH}$ containing only the right-hand part and correspondingly $X^{\rm LH}$ for the left-hand part to define statistics $L_{\rm met}^{\rm RH}$, $L_{\rm met}^{\rm LH}$, $C_{\rm SSM}^{\rm RH}$, $C_{\rm SSM}^{\rm LH}$, etc.
For clarity, we write $L_{\rm met}^{\rm BH}$, $C_{\rm SSM}^{\rm BH}$, etc.\ for statistics calculated for a musical score with both hand parts.
Since the notion of bass notes is not valid for separated hand parts, the log probability of pitch ranks $L_{\rm p.rank}$ is only considered for musical scores with both hand parts.
In total, we have 16 statistics considered for estimating bar line (downbeat) positions: $L_{\rm met}^{\rm BH}$, $L_{\rm met}^{\rm RH}$, $L_{\rm met}^{\rm LH}$, $L_{\rm NV}^{\rm BH}$, $L_{\rm NV}^{\rm RH}$, $L_{\rm NV}^{\rm LH}$, $R^{\rm BH}_{\rm tie}$, $R^{\rm RH}_{\rm tie}$, $R^{\rm LH}_{\rm tie}$, $L_{\rm rel.pc}^{\rm BH}$, $L_{\rm rel.pc}^{\rm RH}$, $L_{\rm rel.pc}^{\rm LH}$, $C_{\rm SSM}^{\rm BH}$, $C_{\rm SSM}^{\rm RH}$, $C_{\rm SSM}^{\rm LH}$, and $L_{\rm p.rank}$.
The statistics are also listed in Table \ref{tab:MathSymbols}.

Most of the statistics formulated in this section involve non-local musical quantities.
For example, even though $L_{\rm met}^{\rm BH}$ is a local statistic as defined in the metrical HMM, $L_{\rm met}^{\rm RH}$ and $L_{\rm met}^{\rm LH}$ are non-local statistics as they involve information of hand parts that is not given a priori (in the transcription task).
To assign a hand part to a note, non-local pitch contexts should be taken into account \cite{Nakamura2014}.
Similarly, the relative pitch class is effectively a non-local quantity as it involves an inference of musical keys that depend on non-local pitch contexts \cite{Krumhansl}.
For inferring note values, it is also necessary to use non-local pitch contexts and inter-dependence of neighbouring quantities \cite{Nakamura2017B}, and thus the related statistics $L^{\rm BH}_{\rm NV}$, $R^{\rm BH}_{\rm tie}$, etc.\ are also non-locally defined.
Statistics based on the SSM are also non-local quantities.

\section{Estimation of Global Characteristics}
\label{sec:EstimationOfGlobalCharacteristics}
%
\begin{figure*}
\centering
\includegraphics[width=1.98\columnwidth]{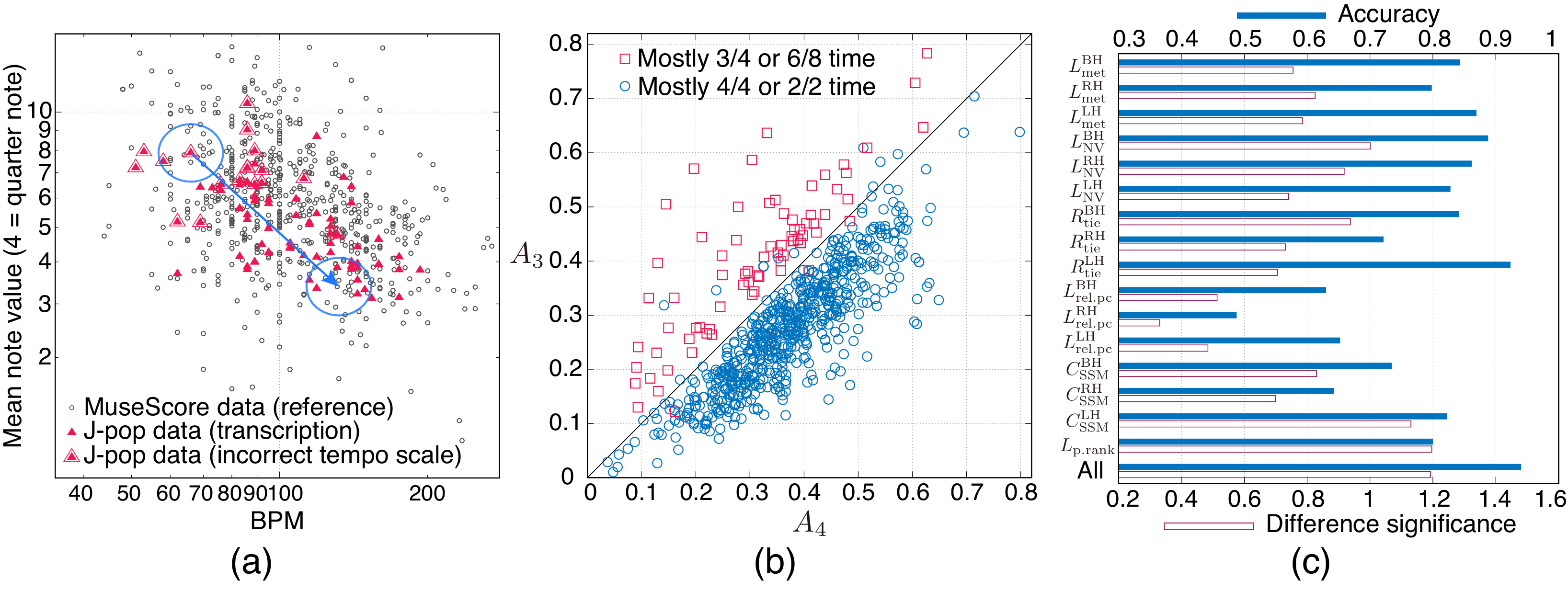}
\caption{Non-local musical statistics and global musical characteristics. (a) Distribution of global tempos represented by beat per minute (BPM) and mean note values. Circles indicate samples in the reference MuseScore data and triangles indicate samples in the scores transcribed from the J-pop data. Transcribed scores with incorrect tempo scales are indicated by outlined triangles. For one such sample, the neighbour is marked by an oval, and similarly for the point obtained by doubling the tempo (indicated by an arrow). (b) Auto-similarity indices for samples in the MuseScore data. (c) Accuracies and difference significances of downbeat estimation on the MuseScore data using each statistic and all the statistics (only pieces in 4/4 time were used). The difference significances are obtained by calculating the differences of the statistics between the original and time-shifted scores and dividing the mean by the standard deviation (averaged for the cases of one, two, and three beat shifts), representing how significant the statistics differ between scores with correct downbeat positions and those with incorrect downbeat positions.}
\label{fig:StatAnalysis}
\end{figure*}

For the non-local nature of the statistics, they cannot be incorporated in the rhythm quantization method in a computationally tractable way.
However, it is possible to utilize them after a preliminary transcription step.
This possibility is suggested by the fact that recognition of onset and offset score times is almost decoupled from recognition of the global characteristics.
We therefore construct post-processing methods for estimating the global characteristics (tempo scale, metre, and bar line positions), using as input a result of transcription by the method of Sec.~\ref{sec:System} (preliminary transcription).
These methods are explained one by one in the following subsections.

\subsection{Tempo Scale Estimation}
\label{sec:TempoScaleEstimation}

The global tempos represented by beat per minute (BPM) and mean note values obtained from the MuseScore data (reference musical scores) and those obtained from the scores transcribed from the J-pop data are plotted in Fig.~\ref{fig:StatAnalysis}(a).
Most samples of the MuseScore data are concentrated in the central region, indicating the convention of tempo scales in the musical genre.
In addition, since the size of the spread of global tempos is comparable to the factor of 2, correct tempo scales cannot be uniquely determined.
We also confirmed that adjusting the prior distribution of the tempo scales (described by $u_{\rm ini}$ and $\sigma_{{\rm ini}\,u}$) to the data distribution did not change the result significantly; it is the likelihood of onset score times that dominantly influences the estimation of tempo scales in the method using the metrical HMM.
For the transcribed scores, pieces with tempo scales different from the ground truth are indicated by outlined triangles.
Most of these cases have tempos smaller than the mean or mean note values larger than the mean, reflecting that most of them have halved tempos.

In this log--log plot, doubling the tempo (and correspondingly halving the note values) can be represented as a translation by a constant distance; an example is indicated by an arrow.
Some transcription samples have a few reference data samples in their neighbours and more of them when their tempos are doubled.
Doubling the tempo of a transcribed score is reasonable in this case according to the knowledge about the data distribution.
Formalizing this idea, we can devise a method for estimating tempo scales: compare the densities of reference data at the original point (transcription score) and the prospective point with a doubled tempo, and if the latter is higher, double the tempo.
We use the kernel density estimation method with a Gaussian kernel in the logarithmic space.

\subsection{Metre Identification}
\label{sec:MeterIdentification}

The auto-similarity indices $A_4$ and $A_3$ for the samples in the MuseScore data are plotted in Fig.~\ref{fig:StatAnalysis}(b).
For this analysis, we selected samples that have a single time signature spanning more than $90\%$ of the durations and used samples with mostly 4/4 or 2/2 time and those with mostly 3/4 or 6/8 time.
It is confirmed that these indices are good discriminators of the metres.
Therefore, a method with a criterion $A_4<A_3$ for identifying a triplet metre can be devised.
The accuracy of the binary classification was $97.8\%$ for these data.

\subsection{Positions of Bar Lines}
\label{sec:DownbeatEstimation}
%
\begin{figure*}
\centering
\includegraphics[width=1.4\columnwidth]{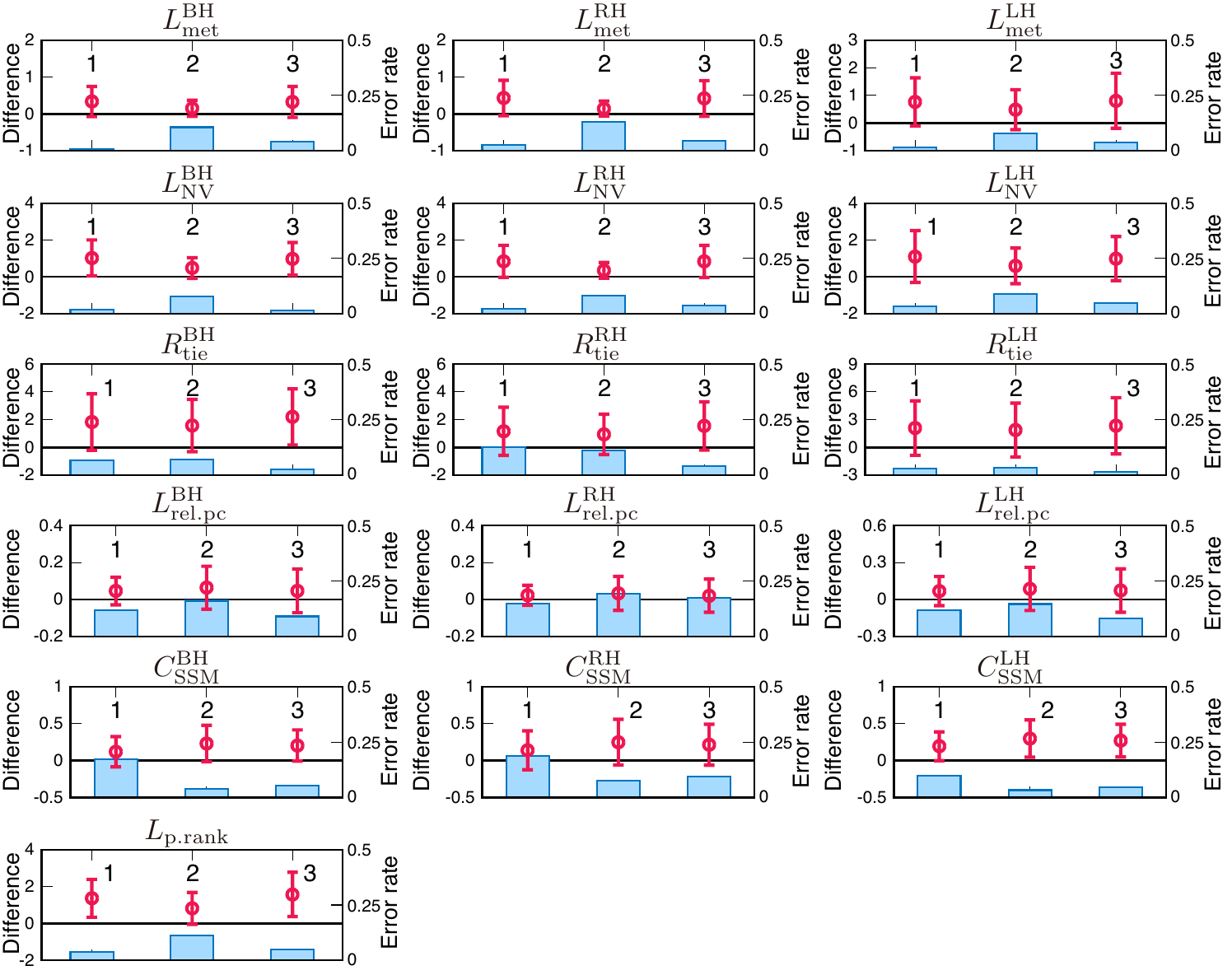}
\caption{Error rates of downbeat estimation on the MuseScore data using each statistic (only pieces in 4/4 time were used). In each panel, the blue boxes under the labels 1, 2, and 3 indicate the frequencies of errors where the estimated downbeats were deviated by one, two, and three beats from the correct positions. The mean and standard deviation of the differences of each statistic between the original and time-shifted scores are shown as red circles and bars.}
\label{fig:DetailedAnalysis}
\end{figure*}

The 16 statistics considered for estimating bar line (downbeat) positions were constructed so that they have larger values for musical scores with correctly positioned downbeats than those with misaligned downbeats.
We can devise a method for estimating downbeat positions based on this property, similarly to the method of maximum likelihood estimation.
To estimate the downbeat positions for a given transcribed score, we calculate the statistics for this score and for those scores obtained by applying time shifts of one, two, and three beats.
The values of the statistics are compared among these four versions of the score and the one with the maximal values is estimated as the correct score.
For a score in 3/4 time, a time shift of three beats does not change downbeat positions and it is necessary to compare only three versions in practice.

For calculating the statistics, we apply the method of \cite{Nakamura2014} for separating hand parts in the preliminary transcription.
We also use an HMM for local key detection to calculate the relative pitch classes, which is a probabilistic variant of the Krumhansl-Schmuckler method \cite{Krumhansl}.
For the calculation of an SSM, we set $\Delta$ to the bar length $M$ of the preliminary transcription result.
For the calculation of pitch rank, we set $K=10$ because there are usually 10 to 20 notes in a bar and a span of each bass note is usually one bar or a half.
Prior to the analysis, the parameters of the log probabilities $L^{\rm BH}_{\rm met}$, $L^{\rm BH}_{\rm NV}$, etc.\ were learned from the MuseScore dataset.

To investigate the effect of each of the 16 statistics, we first analysed the accuracy of downbeat estimation using each statistic alone on the MuseScore data.
We used pieces in 4/4 time (a dominant part of the data) and tested whether the method can correctly reproduce the correct downbeat positions.
Results are shown in Fig.~\ref{fig:StatAnalysis}(c), where the significances of the differences of the statistics between the original and time-shifted scores are also shown.
First, since the chance rate of this estimation problem is $25\%$, every statistic had some positive effect in estimating downbeat positions.
On the other hand, as a single statistic, only $L^{\rm LH}_{\rm met}$, $L^{\rm BH}_{\rm NV}$,  $L^{\rm RH}_{\rm NV}$, and $R^{\rm LH}_{\rm tie}$ had a higher accuracy than $L^{\rm BH}_{\rm met}$, which is equivalent to the metrical HMM and considered as a baseline.
As expected, a statistic with a large difference significance generally had a high accuracy.
A notable exception is $R^{\rm LH}_{\rm tie}$, whose relatively low significance is caused by a large variance of this quantity.

When downbeat positions are incorrectly estimated, they deviate from the correct positions by one, two, or three beats, and the frequencies of these errors are separately shown in Fig.~\ref{fig:DetailedAnalysis}.
For most statistics, deviations of downbeat positions in two beats (or a half bar) were the most frequent errors, which is reasonable given that 4/4 time is a composite metre and both the first and third beats are strong.
For the other statistics, $R^{\rm RH}_{\rm tie}$ and the SSM contrast indices, in contrast, the most frequent errors were deviations in one beat, which is a consequence of anticipations frequently used in popular music.
These results indicate that different statistics capture different musical aspects regarding downbeat positions and suggest that it is effective to use them in combination.

To combine the 16 statistics for downbeat estimation, each statistic is standardized on the MuseScore data to zero mean and unit variance.
We calculate the sum of the standardized statistics for an input score and its time-shifted versions and obtain the one that maximizes the value to estimate downbeat positions.
For the MuseScore data, the accuracy when all the statistics are used is shown in Fig.~\ref{fig:StatAnalysis}(c), which was higher than the best value obtained by any single statistic.
In general, we can optimize the combination of used statistics.
There are $2^{16}=65536$ possible combinations and we notate a particular combination as a binary vector called a {\it criterion vector}.
For example, 100-001-010-000-000-1 means that $L^{\rm BH}_{\rm met}$, $L^{\rm LH}_{\rm NV}$, $R^{\rm RH}_{\rm tie}$, and $L_{\rm p.rank}$ are used (the order of the statistics is shown in Fig.~\ref{fig:CrietrionVector}(b)).

For optimization, we use the J-pop data and the transcribed scores obtained by the method in Sec.~\ref{sec:System}.
Similarly as for 4/4 time, we calculated the statistics for triplet metre using the MuseScore data and used them to obtain the standardized statistics.
We used the separate datasets for optimization and training to avoid overfitting.
We applied the aforementioned methods for tempo scale estimation and metre estimation before the application of the downbeat estimation method using the statistics.

The results are shown in Fig.~\ref{fig:CrietrionVector}.
Compared to the baseline method using only $L^{\rm BH}_{\rm met}$ (equivalent to the metrical HMM), the best criterion vectors improved ${\cal F}_{\rm DB}$ by $11.3$ PP, demonstrating the efficacy of using the non-local statistics.
It was also found that using all statistics is better than the baseline but is not the optimal choice.
To find out the most relevant statistics, we calculated the average usage of each statistic in the top-ranked criterion vectors.
The result in Fig.~\ref{fig:CrietrionVector}(b) shows that highly relevant statistics were $L^{\rm RH}_{\rm NV}$, $L^{\rm LH}_{\rm NV}$, $L^{\rm LH}_{\rm rel.pc}$, and $C^{\rm LH}_{\rm SSM}$.
The relevance of statistics obtained from the left-hand part can be explained by the fact that syncopations are less frequent in the left-hand part than in the right-hand part.
In contrast, $R^{\rm RH}_{\rm tie}$, $R^{\rm LH}_{\rm tie}$, and $L_{\rm p.rank}$ played little roles in the combined estimation.
It is likely that the first two statistics lost relevance due to the presence of more detailed statistics $L^{\rm RH}_{\rm NV}$ and $L^{\rm LH}_{\rm NV}$.
Although we do not have a good explanation for the low relevance of $L_{\rm p.rank}$, it is possible that its effect was shaded by the presence of $L^{\rm LH}_{\rm rel.pc}$ and $C^{\rm LH}_{\rm SSM}$, which also take pitch contents into account.
The statistics used in the three best criterion vectors in Fig.~\ref{fig:CrietrionVector}(a) almost coincide with the statistics with the highest average usage in the top-ranked criterion vectors and can be interpreted similarly.

\subsection{Integrated Method and Final Evaluation}
\label{sec:FinalEvaluation}

\begin{figure}
\centering
\includegraphics[width=0.7\columnwidth]{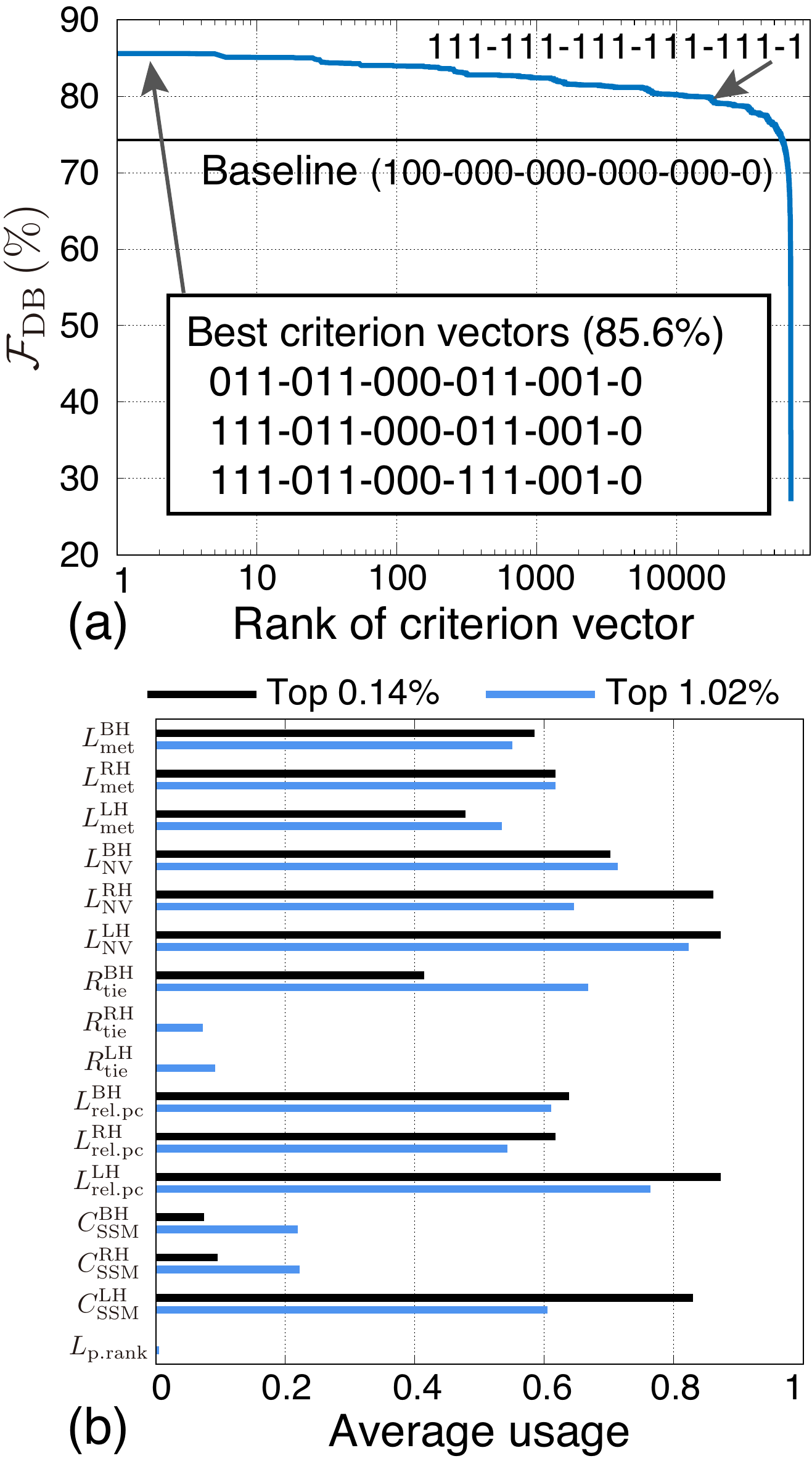}
\caption{Optimization of the criterion vector for downbeat estimation on the J-pop data. (a) Downbeat F-measures for all possible combinations of statistics used for estimation (sorted in F-measures). (b) Average usage of statistics in the top-ranked criterion vectors.}
\label{fig:CrietrionVector}
\end{figure}

Based on the results in the previous subsections, we devised an improved method for piano transcription by integrating the estimations using the non-local statistics.
After a preliminary transcription result is obtained by the method in Sec.~\ref{sec:System}, the method for tempo scale estimation (Sec.~\ref{sec:TempoScaleEstimation}), the method for metre identification (Sec.~\ref{sec:MeterIdentification}), and the method for downbeat estimation (Sec.~\ref{sec:DownbeatEstimation}) are applied sequentially.
A change of tempo scale is applied, if necessary, by multiplying the onset and offset score times by a factor of $2$.
To correct downbeat positions, we shift all the onset and offset score times by one, two, or three beat lengths.
The barline positions are then automatically determined by the identified metre (bar length).
For tempo scale estimation, the standard deviation of the Gaussian kernel was roughly optimized and set to $0.01$.
Since BPMs larger than $200$ are rare, we apply this method only when the transcribed score has a BPM less than $100$.
For downbeat estimation, we use the criterion vector 011-011-000-011-001-0, which is optimal and uses the least number of statistics (Fig.~\ref{fig:CrietrionVector}(a)).
\begin{table}[t]
\centering
\tabcolsep = 3pt
\begin{tabular}{lcccccc}
\toprule
Method & ${\cal E}_{\rm p}$ & ${\cal E}_{\rm m}$ & ${\cal E}_{\rm e}$ & ${\cal E}_{\rm on}$ & ${\cal E}_{\rm off}$ & ${\cal E}_{\rm all}$
\\
\midrule
POVNet+RQ & 0.62 & 4.09 & 7.35 & 2.50 & 20.8 & 7.06
\\
POVNet+RQ+NL & 0.62 & 4.09 & 7.35 & 2.49 & 20.8 & 7.07 
\\
\bottomrule
\toprule
Method & ${\cal F}_{\rm p}$ & ${\cal F}_{\rm voi}$ & ${\cal F}_{\rm met}$ & ${\cal F}_{\rm val}$ & ${\cal F}_{\rm harm}$ & ${\cal F}_{\rm MV2H}$
\\
\midrule
POVNet+RQ & 93.2 & 79.4 & 80.3 & 95.2 & 92.0 & 88.0
\\
POVNet+RQ+NL & 93.2 & 79.7 & {\bf 84.3} & 95.6 & 91.7 & 88.9
\\
\bottomrule
\end{tabular}
\caption{Error rates (\%) and accuracies (\%) of transcription on the J-pop data. `RQ' refers to the rhythm quantization method based on the metrical HMM and `NL' the method using the non-local statistics.}
\label{tab:FinalEvaluationOverall}
\end{table}
\begin{table*}[t]
\centering
\tabcolsep = 5pt
\begin{tabular}{lcccccccc}
\toprule
Method & ${\cal A}_{\rm metre}$ & ${\cal A}_{\rm tempo}$ & ${\cal P}_{\rm B}$ & ${\cal R}_{\rm B}$ & ${\cal F}_{\rm B}$ & ${\cal P}_{\rm DB}$ & ${\cal R}_{\rm DB}$ & ${\cal F}_{\rm DB}$\\
\midrule
POVNet+RQ & 87.7 & 76.5 & 95.1 & 87.1 & 89.8 & 74.9 & 67.1 & 69.4 \\
POVNet+RQ+NL & {\bf 97.5} & {\bf 82.7} & 94.9 & {\bf 90.6} & {\bf 91.9} & {\bf 89.2} & {\bf 84.2} & {\bf 85.6} \\
\bottomrule
\end{tabular}
\caption{Accuracies (\%) of metrical structure. The data and methods are the same as in Table \ref{tab:FinalEvaluationOverall}.}
\label{tab:FinalEvaluationMetrical}
\end{table*}
The final evaluation results for the integrated method are presented in Tables \ref{tab:FinalEvaluationOverall} and \ref{tab:FinalEvaluationMetrical}.
Significant improvements were achieved for the accuracies related to metrical structure and tempo scale.
The metre accuracy was $97.5\%$ (two pieces had incorrect metres), which clearly shows the effect of the auto-similarity measure.
In one case 6/8 time was recognized as 4/4 time, which was partly correct as we can represent a piece in 6/8 time with a 2/4 time signature using triplet notes.
Although the improvement in the tempo scale accuracy indicates that the data distributions of global tempos and mean note values are significant clues for determining tempo scales, intrinsic ambiguities still remained.
For the downbeat F-measure, the significance of using the non-local statistics is evident.
The remaining errors in beat and downbeat positions are caused by misidentifications of metre and tempo scale, deviations of beat times due to transcription errors, and for some pieces, the existence of mixed metres such as an inserted 2/4 bar.
Overall, it has been confirmed that the non-local statistics are useful guides for estimating global musical characteristics in the audio-to-score transcription task.

Examples of transcribed scores are available on the Web\footnote{\url{https://audio2score.github.io/}}.
A music expert may find many unsatisfactory points in these results.
In many cases, bar lines are misaligned at least partly, due to an insertion of an irregular metre or a large tempo change.
Short notes are often deleted in very fast passages (e.g.\ Examples 9 and 11).
There are also many cases of inappropriate voice configurations, which make visual recognition of music difficult.
Despite these limitations, the generated scores can partially be used for music performance and can assist human transcribers, demonstrating the potential of the present method for practical applications.

\section{Discussion}
\label{sec:Discussion}

Here we discuss our results in relation to existing studies and implications for further studies on automatic transcription.
First, estimation of metrical structure has been studied in relation to metre detection \cite{DeHaas2016,Klapuri2003,Temperley1999}, beat tracking \cite{Dixon2000,Krebs2015,Peeters2010}, and rhythm quantization \cite{Foscarin2019,Nakamura2017,Raphael2002}, and the non-local statistics studied here or similar musical features have been considered.
Whereas these studies focused on one or a few of the non-local statistics, they are investigated comprehensively in this study.
An important insight obtained from our result is that, while the statistics work more effectively in combination, using all the statistics is not optimal.
In general, we can introduce real-valued weights for the statistics or use those statistics as inputs to deep neural networks or other classification methods, to further enhance the accuracies.
For these methods to work without overfitting, however, we need much more data.
Another insight is the importance of using statistics based on the separate hand parts.
While the structure consisting of two hand parts is specific to piano music, distinction between low pitch notes (bass and accompaniment parts) and high pitch notes (melody notes) is considered useful for other instrumentations.
Although we focused on the popular music data in the second part of this study, it is expected that the methodology can be applied to music of other genres since the non-local statistics were formulated based on general properties of tonal music \cite{GTTM}.

Second, we found that it is necessary to handle mixed metres (i.e.\ short insertions of irregular metres) for improving the recognition of metrical structure.
Mixed metres are often found in popular music, and fermatas also give a similar effect with regard to rhythm quantization.
Most existing models of musical rhythms assume a fixed metre within a musical piece and a new methodology must be sought to handle this more general case.
As repetitions and other global musical features are considered to be important clues for the recognition of mixed metres, our findings are expected to be useful for solving this problem.

Third, while most previous efforts have been devoted to the improvement of pitch detection and onset rhythm quantization, as reviewed in the Introduction, the final evaluation result in Table \ref{tab:FinalEvaluationOverall} suggests that further investigation is needed for the tasks of note value recognition and voice separation.
The voice separation method devised in this study is based on a hand-crafted cost function, for which precise parameter optimization is difficult, and developing a learning-based method is considered to be an effective approach.
Another possibility is to extend the existing methods assuming monophonic voices \cite{DeValk2018,McLeod2016} to allowing homophonic voices.
Since configurations of note values are closely related to voice structure \cite{Nakamura2017B}, an even promising approach is to jointly estimate them.

Lastly, our results suggest that the following open problems are important in view of practical applications.
To increase the accuracy of rhythm quantization, ornaments such as trills, arpeggios, grace notes, and glissandos should be handled.
To increase the visibility of transcribed scores, clef changes must be placed for pieces with a wide pitch range.
Since the frequency and positions of clef changes are determined by music content and optimized to increase the visibility, this is a non-trivial optimization problem.
Recognition of pedal events, dynamics \cite{Jeong2020}, slurs, articulations, pitch spelling \cite{Bora2019}, and fingering numbers \cite{Nakamura2020} are also necessary to obtain complete musical scores.

\section{Conclusion}
\label{sec:Concl}

In this paper we studied an audio-to-score piano transcription method integrating a DNN-based multipitch detection and statistical-model-based rhythm quantization, and a method for improving the results by using non-local statistics.
In the first part, we confirmed a significant effect of the improved multipitch detection method: on the conventionally used classical music data, the edit-distance-based error rates were reduced by more than half compared to the previous state-of-the-art system \cite{Nakamura2018ICASSP}; and on the popular music data transcribed scores were partly at a practical level.
Transcription errors related to metrical structure were analysed in detail and misidentifications of tempo scale, metre, and positions of bar lines were found to be the most common errors.

In the second part, we studied non-local statistics that serve as guides for recognizing these global musical characteristics.
We found that data distributions of global tempos and mean note values can reduce the ambiguity of tempo scales, that the auto-similarity measures can accurately estimate the metre, and that statistics related to configuration of onset times, note values, relative pitch classes, and the contrast of bar-level SSM were found to be effective for downbeat estimation.
The final evaluation results with the integrated method incorporating these non-local statistics suggested that it is now important to redirect attention to the recognition of note values, voice structure, and other delicate musical score elements that are significant for music performance.

\section{Data Availability}
\label{sec:DataAvailability}

The following contents are available\footnote{\url{https://audio2score.github.io/} (Data.zip)}.
Due to the copyright, it is not permitted to publish the J-pop and MuseScore datasets as well as the transcribed results for these datasets.
However, the lists of URLs where the data were collected are available, by which the datasets can be reproduced.
The transcribed results for the MAPS-ENSTDkCl dataset are available (performance MIDI outputs and MusicXML outputs).
The source code for the multipitch detection method (POVNet), rhythm transcription method, and the method of using non-local statistics is available.

\appendix

\section{Voice Separation Method}
\label{app:VoiceSeparation}

%
\begin{figure}
\centering
\includegraphics[width=1\columnwidth]{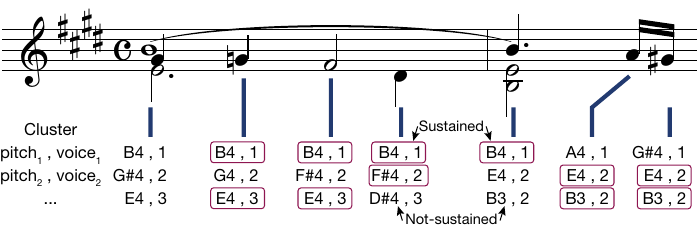}
\caption{Representation of voice configurations.}
\label{fig:voicesep}
\end{figure}
Our method for voice separation is based on sequential optimization using a cost function describing the appropriate structure of voices and the degree of match of this structure to an input quantized MIDI sequence separated into two hand parts.
We apply voice separation for each hand part independently.
The data unit for sequential optimization is a set of notes with simultaneous onset times.
We construct a cluster of these notes for each onset time and we also include in the cluster notes with earlier onset times that temporally overlap with these notes, which we call sustained notes (Fig.~\ref{fig:voicesep}).
We describe voices as integer labels $1,2,\ldots,V_{\rm max}$ given to individual notes of these clusters.
The maximum number of voices $V_{\rm max}$ is a variable that can be set by a user.
For each cluster $C_k$, a set of voice labels $S_k=(s_n)$ for notes $n\in C_k$ in the cluster is called a voice configuration.
The search space for voice separation is the set of all possible voice configurations for all the clusters.

The cost function is constructed as a sum of vertical and horizontal costs defined as follows.
The vertical cost $V(S_k)$ describes the appropriateness of a voice configuration for a cluster and is a sum of four factors:
\vspace{-2pt}
{\setlength{\leftmargini}{13pt}
\begin{itemize}\setlength\itemsep{-0.2em}\setlength{\itemindent}{-0pt}
 \item Assign the value of $s_n$ for each note $n\in C_k$ (penalize unnecessary voices).
 \item Assign $\lambda_2$ for each pair of notes whose voice order and pitch order are opposite (penalize voice crossings).
 \item Assign $\lambda_3$ for each pair of notes having the same voice label but different offset times.
 \item Assign $\lambda_4$ for each pair of sustained and not-sustained notes with the same voice label.
\end{itemize}}
\noindent
The horizontal cost $H(S_{k-1},S_{k})$ describes the appropriate connection of voice configurations of consecutive clusters and is a sum of three factors:
\vspace{-2pt}
{\setlength{\leftmargini}{13pt}
\begin{itemize}\setlength\itemsep{-0.2em}\setlength{\itemindent}{-0pt}
 \item Assign $\lambda_5$ for a sustained note with an inconsistent voice label.
 \item Assign $\lambda_6$ for each pair of consecutive notes with the same voice label having a temporal gap (penalize rests).
 \item Assign $\lambda_7$ for each pair of consecutive notes with the same voice label that temporally overlap.
\end{itemize}}

The sequential optimization can be performed using the Viterbi algorithm.
After the voices are estimated, offset times are corrected  according to the estimated voices, to conform with the constraints that offset times of chordal notes in a voice must match and must be same as or less than the next onset time of that voice.

In the transcription experiments, we fixed a parameterization of the cost function after several trials as $(\lambda_2,\ldots,\lambda_7)=(3,1,1,5,0.2,1)$, and there is room for systematic optimization of the parameters.
We also set $V_{\rm max}=2$ for both hand parts.

\section*{Acknowledgements}

We are grateful to Yu-Te Wu and Li Su for providing the source code of their algorithm in \cite{Wu2019}, Curtis Hawthorne for providing the results obtained by the algorithms in \cite{Hawthorne2018,Hawthorne2019}, and Andrew McLeod for useful discussions and a careful reading of the preliminary version of the manuscript.
We also would like to thank Emmanouil Benetos, Simon Dixon, and Ryo Nishikimi for fruitful discussions.

%
%

\bibliographystyle{apa}

\end{document}